\documentclass[superscriptaddress,aps,onecolumn,12pt,floatfix,altaffilletter,showpacs,showkeys,notitlepage,nofootinbib]{revtex4-2}

\pdfoutput=1

\usepackage[utf8]{inputenc}
\usepackage[dvipdf,dvips,dvipdfmx]{graphicx}
\usepackage[colorlinks=true,citecolor=blue,linkcolor=blue]{hyperref}
\usepackage{amsmath}
\usepackage{amssymb}
\usepackage{soul,xcolor}
\usepackage{epsfig}
\usepackage{epstopdf}
\usepackage{url}
\usepackage{float}
\usepackage{slashed}
\usepackage{cancel}
\usepackage{textcomp}
\usepackage{calc}
\usepackage{adjustbox}
\usepackage{mathtools}
\usepackage{gauss}
\usepackage{physics}
\usepackage{bbm}
\usepackage{esint}
\usepackage{tikz-feynman}
\usepackage{bbold}
\usepackage{tensor}
\usepackage{siunitx}

\usepackage{placeins}
\usepackage{indentfirst}
\usepackage[font=small,labelfont=bf]{caption}
\usepackage{wrapfig}
\usepackage{comment}
\usepackage{pgfplots}
	\pgfplotsset{width=5cm,compat=1.9}
\usepackage{bm}
\usepackage[bottom]{footmisc}
\usepackage{setspace}
\usepackage{fancyhdr}
\setstcolor{red}

\allowdisplaybreaks

\catcode`\@=11
\def\lsim{\mathrel{\mathpalette\@versim<}}
\def\gsim{\mathrel{\mathpalette\@versim>}}
\def\@versim#1#2{\vcenter{\offinterlineskip
\ialign{$\m@th#1\hfil##\hfil$\crcr#2\crcr\sim\crcr } }}
\catcode`\@=12

\newcommand{\met}{g_{\mu\nu}}

\newcommand{\mmet}{\eta_{\mu\nu}}

\newcommand{\grav}{h_{\mu\nu}}
\newcommand{\igrav}{h^{\mu\nu}}
\newcommand{\gdet}{\sqrt{-g}}

\newcommand{\trans}{\enskip\rightarrow\enskip}

\newcommand{\nreturn}{\nonumber \\[0.3em]}
\newcommand{\return}{\\[0.3em]}

\newcommand{\hs}[1]{h_{#1}{}^{#1}}
\newcommand{\hts}[1]{\tilde{h}_{#1}{}^{#1}}

\begin{document}

\title{Inflation with Massive Spin-2 Ghosts}

\author{Jisuke \surname{Kubo}}
\email{kubo@mpi-hd.mpg.de}
\affiliation{Max-Planck-Institut f\"ur Kernphysik (MPIK), Saupfercheckweg 1, 69117 Heidelberg, Germany}
\affiliation{Department of Physics, University of Toyama, 3190 Gofuku, Toyama 930-8555, Japan}

\author{Jeffrey \surname{Kuntz}}
\email{jkuntz@mpi-hd.mpg.de}
\affiliation{Max-Planck-Institut f\"ur Kernphysik (MPIK), Saupfercheckweg 1, 69117 Heidelberg, Germany}

\author{Jonas \surname{Rezacek}}
\email{rezacek@mpi-hd.mpg.de}
\affiliation{Max-Planck-Institut f\"ur Kernphysik (MPIK), Saupfercheckweg 1, 69117 Heidelberg, Germany}

\author{Philipp \surname{Saake}}
\email{saake@mpi-hd.mpg.de}
\affiliation{Max-Planck-Institut f\"ur Kernphysik (MPIK), Saupfercheckweg 1, 69117 Heidelberg, Germany}

\date{\today}

\begin{abstract}
We consider a generic model of quadratic gravity coupled to a single scalar and investigate the effects of gravitational degrees of freedom on inflationary parameters. We find that quantum corrections arising from the massive spin-2 ghost generate significant contributions to the effective inflationary potential and allow for a realization of the spontaneous breakdown of global scale invariance without the need for additional scalar fields. We compute inflationary parameters, compare the resulting predictions to well-known inflationary models, and find that they fit well within the Planck and BICEP/Keck collaboration's constraints on inflation.
\end{abstract}


\maketitle

\pagebreak

\raggedbottom
\section{Introduction}

Cosmic inflation is the most promising solution to many puzzles surrounding the big bang and offers a mechanism to generate cosmological perturbations from primordial quantum fluctuations. The approximate scale invariance of the corresponding inflationary power spectrum reported by the Planck and BICEP/Keck collaborations \cite{Akrami2020,BICEP:2021} may also be hinting towards an underlying theory that is scale invariant before dynamical symmetry breaking; a possibility that we will embrace as others have done in the past \cite{Salvio2014, Salvio:2020axm, Kannike_2015,Farzinnia:2015,Karam2019,Kubo2021,GarciaBellido:2011de,Rinaldi:2015uvu, Ferreira:2016wem, Benisty:2018fja, Barnaveli:2018dxo, Ghilencea:2018thl, Kubo:2018kho, Ishida:2019wkd,Kannike:2014mia,Barrie:2016rnv,Vicentini:2019etr,Gialamas:2020snr,Gialamas:2021enw,Aoki:2021skm}.

Many successful models of inflation are constructed around $f(R)$ gravitational sectors which may generically contain more than one power of the Ricci scalar e.g.\ the  Starobinsky model \cite{Starobinsky:1980te} where the additional degree of freedom (DOF) due to the $R^2$ term plays the role of the inflaton. More general higher order terms built from the other independent contractions of the Riemann tensor are rarely included in the action, however, these terms are necessarily generated by quantum effects even if they are not included at the classical level from the start \cite{tHooft:2011aa}. The inclusion of said higher-order curvature terms is known to have an effect on inflationary predictions. In particular, the Weyl tensor squared term ($C^2$) contributes to primordial tensor fluctuations, implying that the ratio of the amplitude of tensor perturbations to those of scalar perturbations
(tensor-to-scalar ratio $r$) receives a correction \cite{Baumann:2015xxa,Salvio:2017xul,Anselmi2020}. Beyond this fact, we will demonstrate that the presence of the massive spin-2 ghost that originates from the $C^2$ term can also generate an inflationary potential that dynamically induces the Planck scale via radiative corrections \'{a} la Coleman-Weinberg \cite{Coleman1973a}. It should be noted that the scalaron degree of freedom originating from the $R^2$ term alone is not sufficient for triggering radiative symmetry breaking in the Jordan frame. In this respect, our considerations will allow for the construction of the most minimal scale invariant model that yields a dynamical generation of the Planck scale and inflationary potential, as no additional bosonic degrees of freedom besides one external scalar and the metric degrees of freedom are necessary. Though similar minimal-scalar-content models of scale invariant inflation have been studied before \cite{Salvio2014,Salvio:2020axm}, to our knowledge the present work is the first time that they are shown to lead to both a valid inflationary potential and to the spontaneous breaking of scale invariance due to the one-loop contribution of the massive spin-2 ghost in the effective potential, which we compute explicitly in this work.

It is well-known that the massive ghost DOFs that appear when one considers the $C^2$ term in the action threaten the unitarity of the resulting quantum theory. This quantum version of the Ostrogradsky instability, usually referred to as the ``ghost problem'', is a subtle and complicated issue that we will not directly address in this work, as many interesting methods of establishing unitarity in the presence of four derivatives already exist in the literature. The fakeon prescription of Anselmi \cite{Anselmi2017,Anselmi2018} is one such method, where the Lee-Wick approach \cite{Lee1969a,Lee1970} is employed after quantizing the ghost as a ``fake'' degree of freedom that does not appear in the asymptotic spectrum of physical states. There are also several proposed solutions to the ghost problem that leave the ghost's status as a genuine physical particle intact, a view that we will adopt in this work as well. Notable examples in this spirit include the work of Salvio \cite{Salvio2019a,Salvio:2020axm} where a new type of norm between quantum states is defined that leaves all quantum mechanical probabilities positive, the work of Donoghue and Menezes that centers around the decay of the massive ghost \cite{Donoghue2019,Donoghue2021}, as well as the interesting possibility of $\mathcal{P}\mathcal{T}$ quantization championed by Bender and Mannheim \cite{Bender2007,Bender2008}. Though the finer details of the works mentioned above are beyond the scope of the current paper, the relevant takeaway from all of them is that the presence of the spin-2 ghost in quadratic gravity does not immediately render the theory unphysical, and if one includes a $C^2$ term in the gravitational action, the physical effects of the massive ghost on inflationary predictions should not be neglected. 

Inflation has been studied in models that employ the previously mentioned resolutions to the ghost problem in the past. In \cite{Anselmi2020}, it is demonstrated that the fakeon prescription leads to a tensor-to-scalar ratio that is compatible with current BICEP/Keck constraints \cite{BICEP:2021}, though this work is based on the most general form of quadratic gravity paired with a classical Starobinsky inflationary potential that does not employ the global scale-invariance and derived Coleman-Weinberg potential that we will use here. Scale-invariant models whose unitarity rests on the arguments of \cite{Salvio2019a,Salvio:2020axm} lead to the inflationary predictions calculated in \cite{Salvio2017,Salvio2019}. In these works, one finds inflationary parameters that also fit nicely within modern constraints, provided that the spin-2 ghost remains heavy enough to avoid kinetic instabilities\footnote{It has recently been shown that the predictions of \cite{Anselmi2020} and \cite{Salvio2017,Salvio2019} coincide with one another in this regime of metastability \cite{Salvio2022}.}. However, in these studies the inflationary potential does not include one-loop contributions from the ghost as it does in the present work.

We begin our investigations in the next section by establishing the full non-linear action, extracting the gravitational degrees of freedom, and deriving the propagators for said DOFs. We next calculate the Coleman-Weinberg one-loop effective potential in the Jordan frame which includes contributions from the scalars and the spin-2 ghost. Then, using this effective potential, we identify the dynamically generated Planck scale and establish the inflationary potential after transforming to the Einstein frame. Finally, we perform a numerical analysis of the predicted inflationary parameters and end with a discussion of the results.

\section{The model}

We consider the following general action describing globally scale invariant quadratic gravity non-minimally coupled to a single additional matter scalar $S(x)$,
\begin{gather}
S_\text{T} = S_\text{QG} + S_\text{S} \label{ST} \,, \return
S_\text{QG} = \int\dd^4x\gdet\Big(\gamma R^2 - \kappa C_{\mu\nu\rho\sigma}C^{\mu\nu\rho\sigma}\Big) \label{SQG} \,, \return
S_\text{S} = \int\dd^4x\gdet\bigg(\frac{1}{2}\nabla_\mu S\nabla^\mu S - \frac{\beta}{2}S^2R - \frac{\lambda}{4}S^4\bigg) \,, \label{SS}
\end{gather}
where $\gamma$, $\kappa$, $\beta$, and $\lambda$ are arbitrary dimensionless constants. As is standard practice in studies of quadratic gravity, the gravitational part of this action is parameterized in terms of the sum of squares of the Ricci scalar and Weyl tensor, which is equivalent to a general combination of the three independent contractions of the Riemann tensor after neglecting total derivatives \cite{Salvio2018}. The complete action (\ref{ST}) is invariant under infinitesimal local diffeomorphisms as well as the global scale transformations
\begin{align}
&\met \trans \omega^2\met \,,     &&S \trans \omega^{-1}S \,,
\end{align}
where $\omega$ is a constant. The presence of this global symmetry is of particular interest because, as laid out in \cite{Kubo2021}, the scalar $S$ may form a condensate $\langle S \rangle = v_S$ that leads to the spontaneous breakdown of scale invariance and the subsequent generation of an Einstein-Hilbert term and identification of the Planck mass $M_\text{Pl} \propto v_S$.

Since we are interested in the effects of gravitational DOFs on inflation, we separate out the dynamical part of the metric by linearizing the action around flat space with
\begin{align} \label{grav}
\met \trans \mmet + \grav \,,
\end{align}
where $\mmet$ is the Minkowski metric and $\grav(x)$ is a small perturbation. In a similar spirit, we also decompose the scalar in terms of quantum fluctuations $\hat S$ around an approximately constant background field $S_\text{cl}$.
\begin{align} \label{scalarbackground}
S \trans S_\text{cl} + \hat S
\end{align}
For the quadratic fluctuations of the action \eqref{ST} in the fields $h_{\mu\nu}$ and $\hat S$ we then obtain
\begin{align} \label{Squad}
 S_\text{T}^{\text{(quad)}} = &\int\dd^4x\bigg[\frac{\beta}{8}S_\text{cl}^2\Big(\igrav\Box\grav + 2\igrav\partial_\nu\partial^\rho h_{\mu\rho} - \hs{\mu}\Box\hs{\nu} - 2\hs{\mu}\partial_\nu\partial_\rho h^{\nu\rho}\Big) \nreturn
&+ \gamma\Big(\igrav\partial_\mu\partial_\nu\partial_\rho\partial_\sigma h^{\rho\sigma} + \hs{\mu}\Box^2\hs{\nu} + 2\hs{\mu}\Box\partial_\nu\partial_\rho h^{\nu\rho}\Big) \nreturn
&+\frac{\kappa}{6}\Big(\!\!-3\igrav\Box^2\grav - 6\igrav\Box\partial_\nu\partial^\rho h_{\mu\rho} - 2\igrav\partial_\mu\partial_\nu\partial_\rho\partial_\sigma h^{\rho\sigma} + \hs{\mu}\Box^2\hs{\nu} \nreturn
& + 2\hs{\mu}\Box\partial_\nu\partial_\rho h^{\nu\rho}\Big) -\frac{1}{2}  \hat S \beta S_\text{cl} \left( \partial_\nu \partial_\mu + \Box \eta_{\mu\nu} \right)h^{\mu\nu} + \frac{1}{2} \hat S \left( \Box - 3\lambda S_\text{cl}^2 \right)\hat S\bigg] \,,
\end{align}
where $\Box=-\partial_\mu\partial^\mu$. In the equation above we have omitted terms related to the tree-level cosmological constant by the background $S_\text{cl}$ and the tree-level equation of motion for constant $S_\text{cl}$.

We may further separate the gravitational DOFs according to their spin by performing a York decomposition in terms of transverse-traceless tensor modes $\tilde{h}_{\mu\nu}(x)$, transverse vector modes $V(x)$, the scalar trace $\hs{\mu}(x)$, and an additional scalar mode $a(x)$ as 
\begin{gather} \label{York}
h_{\mu\nu} = \tilde{h}_{\mu\nu} + \partial_\mu V_\nu + \partial_\nu V_\mu + \left(\partial_\mu\partial_\nu - \frac{1}{4}\eta_{\mu\nu}\Box\right)a + \frac{1}{4}\eta_{\mu\nu}\hs{\rho} \,, 
\end{gather}
where $\partial^\mu\tilde{h}_{\mu\nu}=\hts{\mu}=0$ and $\partial_\mu V^\mu=0$ \cite{Antoniadis1991}. It is also instructive to redefine the graviton trace in terms of the gauge-invariant scalar quantity $\phi(x)$ \cite{Alvarez-Gaume2016},
\begin{align} \label{scalaron}
\phi = \hs{\mu} - \Box\,a \,,
\end{align}
which may be identified as the well-known ``scalaron'' degree of freedom \cite{Starobinsky:1980te}. After applying these definitions, all of the quadratic terms containing $V_\mu$ and $a$ cancel out leaving us with the simple action,
\begin{align} \label{SYork}
S_\text{T}^{\text{(quad)}} = &\int\dd^4x\left[ \frac{9\gamma}{16} \phi\bigg(\Box^2 - m_\phi^2\Box\bigg)\phi - \frac{\kappa}{2}\delta_{\mu\nu\rho\sigma}\tilde{h}^{\mu\nu}\left(\Box^2 - m_\text{gh}^2\Box\right)\tilde{h}^{\rho\sigma}\right. \nreturn
&\left. - \hat S \left(\frac{3}{4} \beta S_\text{cl} \Box \right) \phi +  \frac{1}{2}\hat S \left( \Box -m_S^2\right)\hat S \right] \,,
\end{align}
where $\delta_{\mu\nu\rho\sigma} = \frac{1}{2}(\eta_{\mu\rho}\eta_{\nu\sigma} + \eta_{\mu\sigma}\eta_{\nu\rho})$ and we have identified the field-dependent masses
\begin{align} \label{massdefs}
&m_\phi^2 = \frac{\beta}{12\gamma}S_\text{cl}^2 \,, &m_S^2 =3 \lambda S_\text{cl}^2 \,, &&m_\text{gh}^2 = \frac{\beta}{4\kappa}S_\text{cl}^2 \,. 
\end{align}
Is it important to recall here that $S_\text{cl}$ is a classical (approximately constant) background field and that the corresponding vacuum expectation value $v_S$ is determined with respect to the one-loop effective potential (cf. eq. \eqref{VEV}) which will be computed in the next section.



\section{Inflation}

\subsection{The effective potential}

We proceed with a computation of the quantum effective potential $U_{\text{eff}}$ including one-loop contributions from the massive spin-0 and spin-2 sectors, each of which may be calculated using standard Coleman-Weinberg (CW) methods i.e.\ by integrating out the fluctuations $\psi =(\tilde{h}_{\mu\nu} , \phi , \hat S)^T$ \cite{Coleman1973a}. The part of the functional integral that is quadratic in $\psi$ is Gaussian, leading to an effective potential that is proportional to 
\begin{align} \label{effpot}
U_\text{(1-loop)}(S) &= -\frac{i}{2}\ln\bigg[\text{Det}\bigg(\frac{\delta^2 S_\text{T}^{\text{(quad)}}}{\delta \psi \delta \psi}\bigg)\bigg] \nreturn
&= -\frac{i}{2}\ln(\text{Det}M ) -\frac{i}{2} \Tr[\ln(\delta_{\mu\nu\rho\sigma}\left(-\Box^2 + m_\text{gh}^2 \Box \right))] \,,
\end{align}
where the determinant is taken in the functional sense and from here on we drop the subscript of $S_\text{cl}$ for convenience. 

The first term above corresponds to the scalar DOFs with the Hessian matrix defined as
\begin{align} \label{Mmatrix}
M = \begin{pmatrix}
\frac{9\gamma}{8} \left(\Box^2 - m_\phi^2\Box\right) && -\frac{3}{4} \beta S \Box \\
-\frac{3}{4} \beta S \Box && \Box - m_S^2
\end{pmatrix} \,.
\end{align}
The $\log$ can be rewritten as
\begin{align} \label{scalarDOF}
\ln(\text{Det}M ) =  \Tr[\ln(\Box-m_+^2)] + \Tr[\ln(\Box-m_-^2)] + \cdots
\end{align}
where the ``$\cdots$'' stand for irrelevant constant terms that are independent of $S$ and we have introduced the masses
\begin{align} \label{scalarmasses}
m_\pm^2 = \frac{1}{2}\left(m_S^2 + (1+6 \beta)m_\phi^2\right) \pm \frac{1}{2}\sqrt{\left(m_S^2 + (1+6 \beta)m_\phi^2\right)^2 - 4m_S^2 m_\phi^2} \,,
\end{align}
which agrees with the two mass eigenstates in the Einstein frame found in \cite{Salvio2014}. The trace in \eqref{scalarDOF} may be written as a sum of the momentum space eigenvalues of the operators $\ln(\Box - m_\pm)$ and evaluated using dimensional regularization to give the scalar's one-loop contribution to the effective potential. 
\begin{align}
\label{Uscalar}
U_\text{scal}(S) &= -\frac{i}{2} \sum_{j=\pm} \int\frac{\dd^4p}{(2\pi)^4}\ln\bigg(p^2 - m_j^2\bigg) = \sum_{j=\pm} \frac{1}{64\pi^2}m_j^4\bigg[\ln\bigg(\frac{m_j^2}{\mu^2}\bigg) - \frac{3}{2}\bigg]
\end{align}
Here, we have employed $\overline{\text{MS}}$, introducing the renormalization scale $\mu$ in the process, and we have absorbed the divergent terms into the renormalized constant $\lambda$.

Calculation of the spin-2 part follows in much the same way as the spin-0, with the non-zero contributions coming from the last term in \eqref{effpot}, which can be rewritten as
\begin{align} \label{ghosttrace}
 \Tr[\ln(\delta_{\mu\nu\rho\sigma}\big(-\Box^2 + m_\text{gh}^2 \Box \big))] =  \Tr[\ln(\delta_{\mu\nu\rho\sigma}\big(\Box - m_\text{gh}^2 \big))] + \Tr \ln(-\Box) \,,
\end{align}	
where only the massive part of the inverse propagator contributes.\footnote{Here we dropped the contribution $-\frac{i}{2} \Tr \ln(-\Box)$ to the effective potential which is independent of $S$. Therefore, the overall sign of the ghost contribution is the same as for a normal particle, which is consistent with the $\beta$-function of the quartic coupling (see e.g.\ the $\beta$-functions in \cite{Salvio2014}).} However, when going to momentum space, we must take advantage of the transverse-traceless nature of $\tilde{h}_{\mu\nu}$ to write
\begin{align}
\tilde{h}^{\mu\nu}\delta_{\mu\nu\rho\sigma}\tilde{h}^{\rho\sigma} = \tilde{h}^{\mu\nu}P^{(2)}_{\mu\nu\rho\sigma}\tilde{h}^{\rho\sigma} \,,
\end{align}
where
\begin{align}
P^{(2)}_{\mu\nu\rho\sigma} = \frac{1}{2}\big(\theta_{\mu\rho}\theta_{\nu\sigma} + \theta_{\mu\sigma}\theta_{\nu\rho}\big) - \frac{1}{d-1}\theta_{\mu\nu}\theta_{\rho\sigma} \qquad\text{with}\qquad \theta_{\mu\nu} = \mmet - \frac{p_\mu p_\nu}{p^2} \,,
\end{align}
is a spin-2 projection operator \cite{VanNieuwenhuizen1973}. Making this replacement ensures that we count the correct number of degrees of freedom, which is five for a massive spin-2 field in four dimensions, after noting that
\begin{align}
\text{Tr}\big(P^{(2)}_{\mu\nu\rho\sigma}\big) = \delta^{\mu\nu\rho\sigma}P^{(2)}_{\mu\nu\rho\sigma} = \frac{1}{2}(d + 1)(d - 2) \,.
\end{align}
With these considerations, we find that the massive spin-2 contributes
\begin{align} \label{ghostcontribution}
U_h(S) &= -\frac{i}{2}\lim_{d \rightarrow 4}\bigg[\mu^{4-d} \int\frac{\dd^dp}{(2\pi)^d}\frac{1}{2}(d + 1)(d - 2)\ln\bigg(\frac{p^2 - m_\text{gh}^2}{p^2}\bigg)\bigg] \nonumber \\
&= \frac{5}{64\pi^2}m_\text{gh}^4\bigg[\ln\bigg(\frac{m_\text{gh}^2}{\mu^2}\bigg) - \frac{1}{10}\bigg]
\end{align}
to the effective potential, where we have subtracted the divergent part according to the $\overline{\text{MS}}$ scheme.

Finally, the entire effective potential is given by
\begin{align} \label{Ueff3}
U_{\text{eff}}(S) = \frac{\lambda}{4} S^4 + U_\text{scal}(S) + U_h(S) + U_0 \,,
\end{align}
where $U_0$ is an arbitrary constant background that may be tuned in order to ensure that the classical zero-point energy vanishes, provided that scale invariance is broken spontaneously, which, as we will see in the next section, is indeed the case here. It should also be noted that once the cosmological constant is canceled in this way in the Jordan frame, it remains zero after transforming to the Einstein frame as well.

\subsection{The inflationary action}
The effective action for inflation may be written as
\begin{align} \label{Seff}
S_{\text{eff}} = \int\dd^4x\gdet\bigg(\frac{1}{2}S \Box S - \frac{\beta}{2}S^2R + \gamma R^2 - \kappa C_{\mu\nu\rho\sigma}C^{\mu\nu\rho\sigma} - U_{\text{eff}}(S)\bigg) \,,
\end{align}
where $U_{\text{eff}}$ is the quantum effective one-loop potential we computed in the previous section. The effective one-loop potential \eqref{Ueff3} may be written as \cite{Coleman1973a}
\begin{align} \label{Ueff0}
U_{\text{eff}}(S) = U_0 + \bigg[C_1 + C_2 \ln\bigg(\frac{S^2}{\mu^2}\bigg)\bigg]S^4 \,,
\end{align}
where $C_1$ and $C_2$ are dimensionless constants that depend only on the coupling constants $\lambda$, $\beta$, $\gamma$ and $\kappa$. For the sake of brevity, we do not show the explicit form of $C_1$ and $C_2$ here as they may easily be computed from the contributions to (\ref{Ueff3}).

We may now solve for the vacuum expectation value (VEV) of $S$, $v_S$, which is defined as the minimum of this potential
\begin{align} \label{VEV}
&\frac{\partial U_{\text{eff}}(S)}{\partial S} \, \bigg\rvert_{S=v_S} = 0 \,, &&v_S = \mu \exp\bigg(\!\!-\frac{1}{4} - \frac{C_1}{2C_2}\bigg) \,.
\end{align}
The non-zero value of this minimum indicates a spontaneous breakdown of global scale symmetry. We may also easily calculate the explicit value of $U_0$ by requiring that the effective potential vanishes in the broken phase which yields
\begin{align} \label{CC}
&U_{\text{eff}}(v_S)=0\,, &&U_0 = \frac{\mu^4}{2}C_2 \exp \bigg(\!\!-1 - \frac{2C_1}{C_2}\bigg) \,.
\end{align}
Finally, we obtain the explicit value of the Planck mass that is generated by the breaking of scale invariance by identifying the canonical Einstein term in (\ref{Seff}) as
\begin{align} \label{PlanckIdent}
&-\frac{1}{2}\beta S^2 R \, \bigg\rvert_{S=v_S} = -\frac{1}{2} M_\text{Pl}^2 R \,, &&M_{\text{Pl}}^2 = \beta v_S^2 \,.
\end{align}
In analogy to \cite{Kubo2021}, this relates $M_\text{Pl}$ to the renormalization scale $\mu$ via (\ref{VEV}). In contrast to the aforementioned work \cite{Kubo2021}, we do not need two external scalars to achieve successful Coleman-Weinberg symmetry breaking of scale invariance in the Jordan frame. This is in particular due to the additional contributions from the massive spin-2 ghost to the effective potential which, to our knowledge, is a novel consideration. This is opposed to a more general derivation of this phenomenon at the level of the renormalization group equations as in \cite{Salvio2014, Salvio:2020axm}\footnote{In addition, \cite{Salvio2014, Salvio:2020axm} simultaneously account for a vanishing cosmological constant. We do not account for a dynamical solution to the cosmological constant problem in this work by simply tuning $U_0$ in \eqref{Ueff3}.}.

To calculate predictions of the inflationary parameters that result from spontaneous symmetry breaking, we follow the procedure outlined in \cite{Kubo2021}. Since we have shown that scale invariance is spontaneously broken, we may introduce an auxiliary field to remove the $R^2$ term, then transform to the Einstein frame with a Weyl rescaling. There we find two dynamical scalar fields, $S$ and the scalaron. The corresponding potential exhibits a valley structure \cite{Kubo2021}, a flat direction with steep perpendicular potential lines, and the fields will thus always fall into a trajectory along that flat direction. After solving the minimum equations for the scalaron\footnote{Depending on the parameter configuration, solving for $S$ rather than the scalaron may result in a better description of the flat direction. However, in our case the choice of contour has only a minor influence on the inflationary parameter predictions as both contours are valid for all calculated points. For an extended discussion we refer the reader to section 4.2 and Appendix A of \cite{Kubo2021}.},
the final inflaton potential can be rewritten to depend only on the original external scalar $S$ and the coupling constants $\lambda$, $\beta$, $\gamma$, and $\kappa$.
This effective inflationary action in the Einstein frame has the form 
\begin{align}
S_\text{inf}^\text{E} = \int\dd^4x \gdet \bigg(\!\!-\frac{1}{2} M_\text{Pl}^2 R  - \kappa C_{\mu\nu\rho\sigma}C^{\mu\nu\rho\sigma} + \frac{1}{2} F(S)^2 S \Box S - U_\text{inf}(S)\bigg) \, ,
\end{align}
where $F(S)$ denotes the modification to the kinetic term for $S$ and is given by
\begin{align} \label{field_norm}
F(S) &= \frac{1}{\big(1 + 4A\big)B}\bigg[\big(1 + 4A\big)B + \frac{3}{2}M_\text{Pl}^2\Big(\big(1 + 4A\big)B' + 4A'B\Big)^2\bigg]^{1/2} \,,
\end{align}
where $A$ and $B$ are functions of the scalar field $S$ given by
\begin{align}
&A(S) =\frac{4 \gamma \, U_\mathrm{inf}(S)}{B(S)^2 M_\mathrm{Pl}^4} \,,  &&B(S) = \frac{\beta S^2}{M_\text{Pl}^2} \,,
\end{align}
and primes denote derivatives with respect to $S$. With these definitions, the full inflaton potential $U_\text{inf}(S)$ is thus determined to be \pagebreak
\begin{align} \label{Uinf}
U_\text{inf} (S) =  \frac{U_\text{eff}(S)}{B(S)^2 + 16 \gamma \, U_\text{eff}(S)/ M_\text{Pl}^4} \,.
\end{align}
One may also obtain the canonically normalized field $\tilde{S}$ via a simple integration.
\begin{align} \label{normRelation}
\tilde{S}(S) = \int_{v_S}^S \dd x \, F(x)
\end{align}

\subsection{Numerical analysis of slow-roll inflation} \label{sec:numerical}

Inflationary CMB observables, namely, the scalar spectral index $n_S$ and the tensor-to-scalar ratio $r$, may be expressed in terms of the slow-roll parameters $\varepsilon$ and $\eta$ as
\begin{align}
&n_S = 1 + 2\eta_* - 6\varepsilon_* \,, &&r = 16\varepsilon_* \,,
\end{align}
where we have neglected contributions from the Weyl tensor (which will be addressed later) and the asterisks indicate quantities evaluated at $S=S_*$, the value of S at time of CMB horizon exit. Since our inflaton potential depends only on the scalar field $S$, we can apply the well known formulas for  $\epsilon$, $\eta$, and $N_\text{e}$ of one-field slow-roll inflation, modified to depend on the non-normalized field $S$ using the relation (\ref{normRelation}).
\begin{align}
&\varepsilon (S) = \frac{M_{\rm Pl}^2}{2\,F^2(S)}\left(\frac{U'_{\rm inf}(S)}{U_{\rm inf}(S)}\right)^2 \label{epsilon} \return 
&\eta(S) = \frac{M_{\rm Pl}^2}{F^2(S)} \left(\frac{U''_{\rm inf}(S)}{U_{\rm inf}(S)} -\frac{ F'(S)}{F(S)}\frac{U'_{\rm inf}(S)}{U_{\rm inf}(S)}\right) \label{eta} \return
&N_e = \int_{S_*}^{S_\mathrm{end}}\frac{F^2(S)}{M_{\rm Pl}^2}\frac{U_{\rm inf}(S)} {U'_{\rm inf}(S)} \label{efolding}
\end{align}
\begin{figure}[p]
\centering
\includegraphics[width=0.81\textwidth]{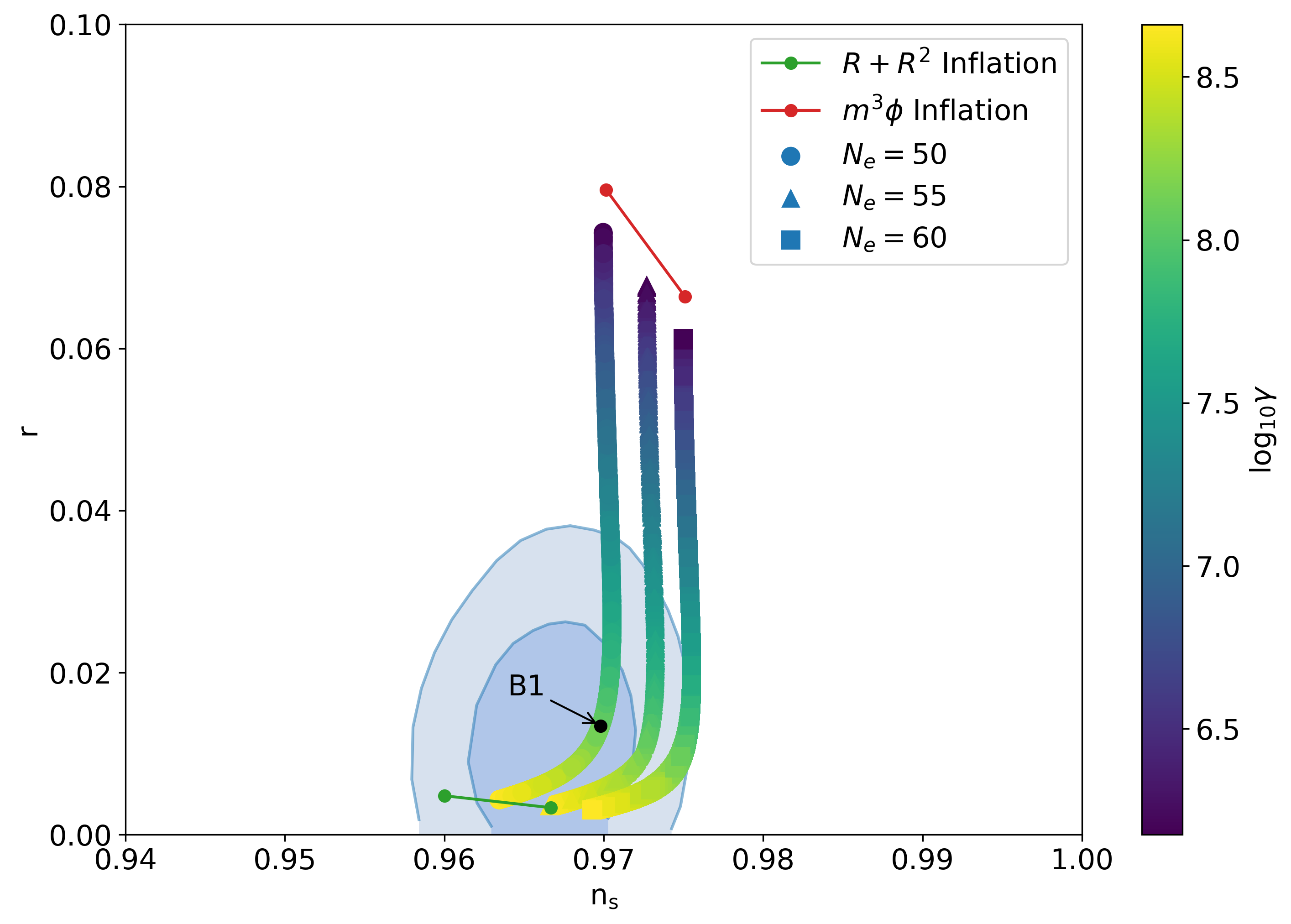}\\
\vspace{0.2cm} \noindent
\includegraphics[width=0.81\textwidth]{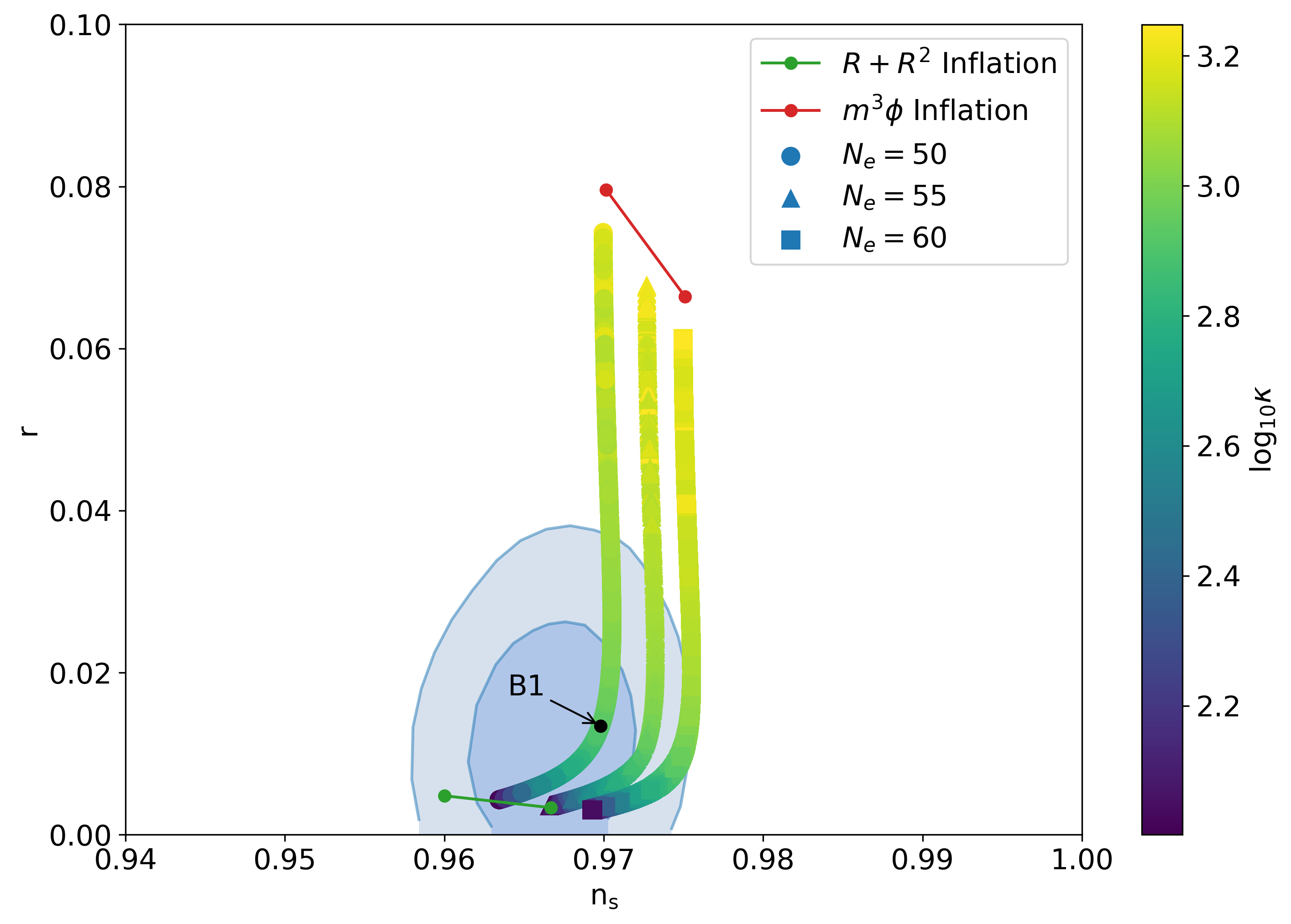}
		\caption{Predictions for the scalar spectral index $n_\mathrm{s}$ and the tensor-to-scalar ratio $r$ with varying numbers of e-folds $N_e$ are displayed. For the points shown, $\lambda$ is fixed, while $\beta, \gamma$, and $\kappa$ are taken randomly from (\ref{couplingvals}) while satisfying (\ref{AsConstraint}). The color gradient shows $\gamma$ in the upper and $\kappa$ in the lower plot. We include the Planck TT,TE,EE+lowE+lensing+BK18+BAO $68\%$ and $95\%$ CL regions from \cite{BICEP:2021,Akrami2020}, as well as predictions of the Starobinsky model (green) and linear inflation (red). Corrections to $r$ due to the $C^2$ term \cite{Baumann:2015xxa,Salvio:2017xul,Anselmi2020} have not been included here, though they are accounted for in Fig.~\ref{correctedPredictions}.}
		\label{ns-r-prediction}
\end{figure}
Here, $S_\text{end}$ denotes the value of $S$ at the end of inflation which is defined by $\text{max}$ $\{\epsilon(S=S_\text{end}),$ $\rvert\eta(S=S_\text{end})\rvert\}=1$. With this we may calculate expressions of $n_S$ and $r$ that depend only on the dimensionless couplings $\lambda$, $\beta$, $\gamma$ and $\kappa$, as $\mu$ is fixed after demanding the correct value for $M_\text{Pl}$ as in (\ref{PlanckIdent}). To constrain this model we use the latest data from the Planck satellite mission \cite{Akrami2020} and assume an inflation duration of $N_e \approx 50-60$ e-folds. To ensure our predictions are consistent with the Planck data, we constrain the parameter space of the dimensionless couplings so that it ultimately fulfills the scalar power spectrum amplitude $A_s$ constraints below.
\begin{align} \label{AsConstraint}
&\ln (10^{10} A_s) = 3.044 \pm 0.014 &&A_s = \frac{U_{\text{inf}\, *}}{24 \pi^2 \epsilon_* M^4_\text{Pl}}
\end{align}
Predictions corresponding to the resulting coupling values below are displayed in Fig.~\!\ref{ns-r-prediction}.
\begin{align} \label{couplingvals}
&\lambda = 0.005 &\beta \in [10^3,10^4] &&\gamma \in [10^3,10^9] &&\kappa \in [10^2,10^{3.25}]
\end{align}
These ranges for the dimensionless couplings\footnote{This choice of parameters corresponds to perturbative values where the logarithms, e.g. $\ln (S_*/v_S)$, do not grow too large during inflation. This ensures that the one-loop CW potential remains a good approximation during inflation and implies that additional RG-running effects are negligible.} are selected after incorporating the Planck constraint on $A_s$ (\ref{AsConstraint}). Though more parameter space was explored, it did not yield promising predictions while remaining compatible with this constraint. To highlight the effect of the $R^2$ and $C^2$ term on the $n_\mathrm{s}-r$ predictions, we display values for their coupling constants ($\gamma$ and $\kappa$) separately as color gradients in the upper ($\gamma$) and lower ($\kappa$) plot of Fig.~\!\ref{ns-r-prediction}. We see that for the full range of possible e-folds, there are points that are compatible with even the tightest Planck and BICEP/Keck constraints \cite{Akrami2020,BICEP:2021}. We also find that the upper limit of our predictions for $r$ approaches the upper limits of linear inflation ($m^3\phi$), while the lower limits match those of Starobinsky inflation. The circles for linear and Starobinsky inflation in Fig.~\!\ref{ns-r-prediction} represent the predictions for $N_e=50$ (left) and $N_e=60$ (right) e-folds respectively. The point labeled ``B$1$'' in Fig.~\!\ref{ns-r-prediction} corresponds to the following benchmark values.
\begin{align} \label{B1}
&\mathrm{B}1: \; \lambda = \SI{0.005}{}  &\beta=\SI{5.62e2}{} &&\gamma= \SI{1.22e8}{} &&\kappa= \SI{837}{} 
\end{align}
In order to get an order of magnitude estimate, we calculate the field masses $m_\phi$ and $m_\text{gh}$ via the relation (\ref{massdefs}) evaluated at the non-zero VEV of $S$,
\begin{align} \label{BPmasses}
&m^{\mathrm{B}1}_\phi(S=v_S) \simeq \SI{6.35e13}{\giga\electronvolt} \,, &&m^{\mathrm{B}1}_\text{gh}(S=v_S) \simeq \SI{4.21e16}{\giga\electronvolt}  \,.
\end{align}
These masses are representative for most of the displayed points, with the field masses of all points displayed in Fig.~\!\!\ref{ns-r-prediction} being roughly contained in the ranges $m_\phi \in \left[ 10^{13} \, \SI{}{\giga\electronvolt} , 10^{14}\, \SI{}{\giga\electronvolt}\right]$ and $m_\text{gh} \in \left[ 10^{16} \,  \SI{}{\giga\electronvolt}, 10^{17} \, \SI{}{\giga\electronvolt} \right]$. Here, low $m_\phi$ goes hand in hand with high $\gamma$ and therefore leads to relatively small tensor-to-scalar ratios $r$ (see upper panel in Fig.~\!\ref{ns-r-prediction}). Thus, for small tensor-to-scalar ratios, the contribution of the scalaron mass $m_\phi$ to the effective potential is relatively small, while on the other hand, the ghost mass $m_\text{gh}^2 \propto 1/\kappa$ can be larger for low $r$ (see lower panel in Fig.~\!\ref{ns-r-prediction}). Therefore, the ghost contribution to the effective potential \eqref{ghostcontribution} is crucial for triggering symmetry breaking while maintaining low $r$, which is a novel consideration with regard to scale invariant models.

As previously mentioned, the Weyl tensor-squared term can contribute non-trivially to tensor perturbations around a de Sitter background, leading to a correction of the tensor-to-scalar ratio \cite{Baumann:2015xxa,Salvio:2017xul,Anselmi2020}. To calculate this correction factor we use the slow-roll approximation during inflation\footnote{To ensure a field value that is representative for inflation we choose $S=S_*$ to calculate $U_\text{inf}(S)$ and $m_\text{gh}(S)$.}, $H^2 \approx U_\text{inf}/(3 M_\text{Pl}^2)$. Using the further developed version of the equation for the correction, i.e.\ (6.51) in \cite{Salvio:2020axm}, we arrive at
\begin{figure}[t]%
\centering
\includegraphics[width=0.81\textwidth]{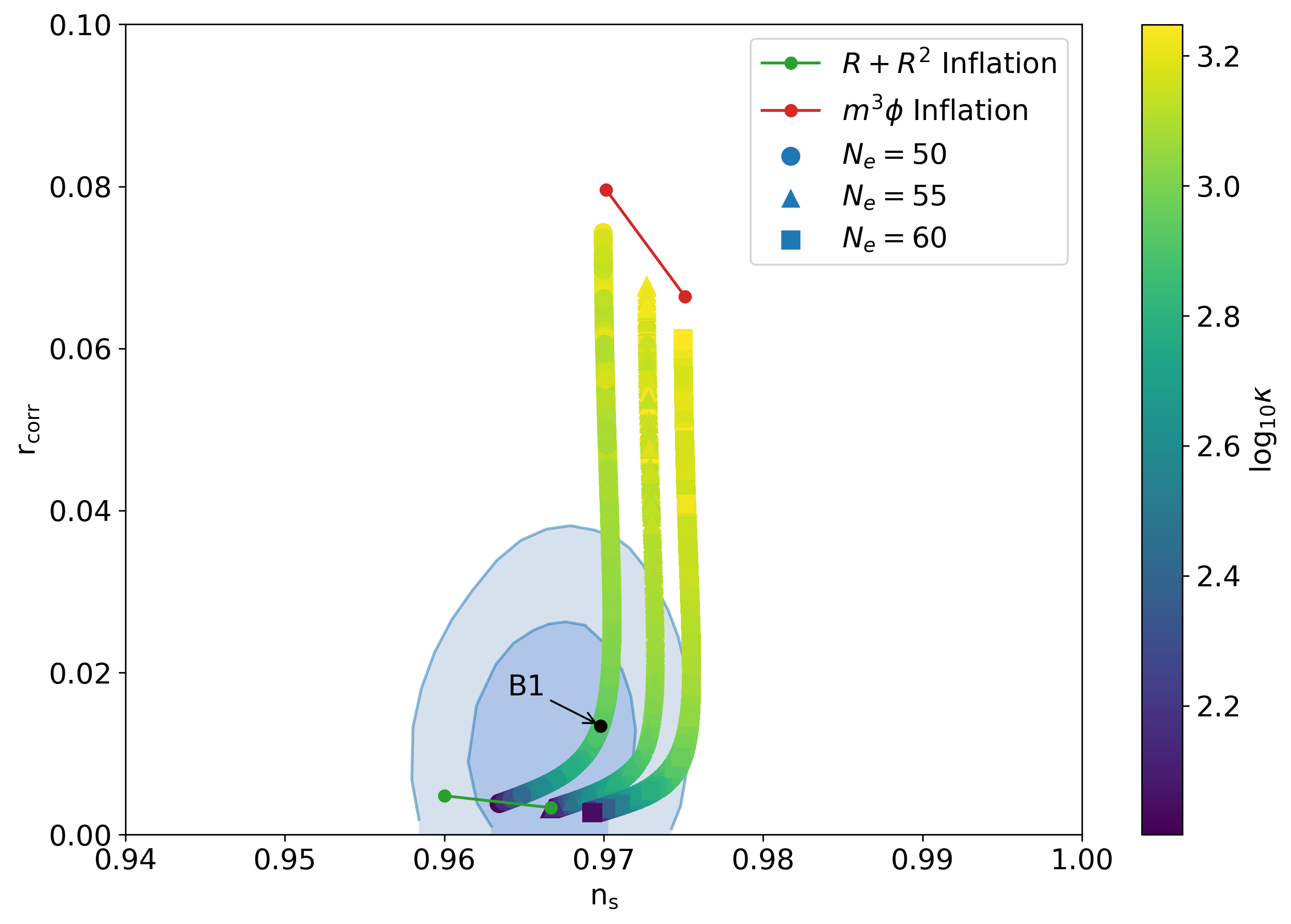}
\caption{Predictions for the scalar spectral index $n_\mathrm{s}$ and tensor-to-scalar ratio $r_\text{corr}$ for different $N_e$.  This includes corrections due to the $C^2$ term given in (\ref{rCorrSalvio}). The data is displayed in the same way as the lower plot of Fig.~\ref{ns-r-prediction}.}
\label{correctedPredictions}
\end{figure}%
\begin{align}
r_\text{corr} &= \; r \left( 1+ \frac{2 \: H^2}{m_\text{gh}^2} \right)^{-1} \simeq \; r \left( 1+  \frac{2 \: U_\text{inf}(S_*)}{3 \: M_\text{Pl}^2 \: m_\text{gh}^2(S_*)} \right)^{-1} \,.
\label{rCorrSalvio}
\end{align}
We find that in the region of interest, i.e.\ small $\kappa$ and large $\gamma$, the tensor-to-scalar ratio is suppressed with a maximum effect of $\approx 11 \%$. Therefore, the corrected predictions are also fully compatible with the strongest cosmological constraints from Planck and BICEP/Keck, as seen in Fig.~\!\ref{correctedPredictions}.

\section{Conclusion}
We have investigated a classical scale invariant framework that dynamically generates the Planck mass via the spontaneous breaking of scale symmetry in the Jordan frame in the most minimal way i.e.\ with only one external scalar in addition to the quantum contributions of the graviton degrees of freedom. Given that higher powers of curvature tensors are necessarily generated via quantum corrections even if these terms are not considered at tree-level, we have included the Weyl tensor squared term from the start. We then find that the resulting quantum contributions allow for spontaneous symmetry breaking via the Coleman-Weinberg mechanism in the Jordan frame. Specifically, it is the massive spin-2 ghost DOF originating from the Weyl squared term that allows for this breaking, a role that is usually filled by additional external scalars in other scale invariant models that also address inflation. The resulting potential in our framework is in perfect agreement with the current strongest constraints from the Planck and BICEP/Keck collaborations, as seen in Fig.~\!\ref{ns-r-prediction} and \ref{correctedPredictions}. Again we stress that the ghost contribution to the effective potential allows for successful dynamical generation of $M_\text{pl}$ while maintaining an inflaton potential that leads to an (observationally favored) low tensor-to-scalar ratio $r$. This feature is not possible if one considers the scalaron DOF alone because of its relatively small contribution to the one-loop effective potential at low $r$. We have also found that this finding is not qualitatively modified if the corrections to the tensor-to-scalar ratio due to the presence of the $C^2$-term calculated in \cite{Baumann:2015xxa,Salvio:2017xul,Anselmi2020} are included (see Fig.~\!\ref{correctedPredictions}). 

To conclude, we note that though the primordial non-Gaussianities in cosmological fluctuations are suppressed in single-field systems of inflation \cite{Maldacena:2002vr}, they can still be generated in a multi-field system and appear in the CMB anisotropy, as well as in measurements of the large scale structure of the Universe (see for instance \cite{Bartolo:2004if} and \cite{Planck:2019kim}). Future experimental projects such as LiteBird \cite{LiteBIRD:2022}, Euclid \cite{Amendola:2016saw}, LSST  \cite{LSSTScience:2009jmu}, etc.\ will be able to measure the magnitude of these non-Gaussianities and constrain their existence. Though our model contains only one non-gravitational scalar field at the beginning, the scalaron which originates from the $R^2$ term in the action (\ref{SQG}) makes the system behave as an effectively two-field system. In \cite{Mori:2017caa} it is outlined how one may compute the non-Gaussianities in such models. Furthermore, as we have mentioned with reference to \cite{Salvio:2017xul} in Section \ref{sec:numerical}, the massive spin-2 mode can contribute to the metric perturbations during inflation, thus altering the inflationary parameters. This correction has turned out to be very small in our model, but the size of its contribution to the non-Gaussianities is not known as of yet. We plan to put the focus of our future investigations on calculating the primordial non-Gaussianities in inflationary models based on scale invariance.


\acknowledgments
We thank Manfred Lindner and Andreas Trautner for many fruitful discussions. J.R.\ and P.S.\ are supported by the IMPRS-PTFS.
J.\ Kubo is partially supported by the Grant-in-Aid for Scientific Research (C) from the Japan Society for Promotion of Science (Grant No.19K03844).


\bibliographystyle{JHEP}
\bibliography{library_new}

\end{document}